\RequirePackage{fix-cm}
\documentclass[twocolumn,epjc3]{svjour3}  
\smartqed  
\RequirePackage{graphicx}
\journalname{Eur. Phys. J. C}
\begin{document}
%
\title{Understanding NaI(Tl) crystal background for dark matter searches}



\author{G.~Adhikari\thanksref{addr2}
        \and  
        P.~Adhikari\thanksref{addr2}
	\and 
	C.~Ha\thanksref{addr1}
	\and 
	E.J.~Jeon\thanksref{corrauthor,addr1}
	\and 
	N.Y.~Kim\thanksref{addr1}
	\and 
	Y.D.~Kim\thanksref{addr1,addr2}
	\and 
	S.Y.~Kong\thanksref{addr2}
	\and 
	H.S.~Lee\thanksref{addr1}
	\and 
	S.Y.~Oh\thanksref{addr2}
	\and 
	J.S.~Park\thanksref{addr1}
	\and 
	K.S.~Park\thanksref{addr1} 
}

\thankstext{corrauthor}{Corresponding author: ejjeon@ibs.re.kr}


\institute{Center for Underground Physics, Institute for Basic Science (IBS), Daejeon 34047, Republic of Korea \label{addr1}
           \and
           Department of Physics and Astronomy, Sejong University, Seoul 05006, Korea \label{addr2}
}

\date{Received: date / Accepted: date}

\maketitle

\begin{abstract}
We have developed ultra-low-background NaI-(Tl) crystals to reproduce the DAMA results
with the ultimate goal of achieving purity levels that are comparable to or better than those of the DAMA/LIBRA crystals. 
Even though the achieved background level does not approach that of DAMA/LIBRA, it is crucial to have a quantitative understanding of the backgrounds.
We have studied background simulations toward a deeper understanding of the backgrounds and developed background models for a 9.16-kg NaI(Tl) crystal used in the test arrangement.   
In this paper we describe the contributions of background sources quantitatively by performing Geant4 Monte Carlo simulations that are fitted to the measured data to quantify the unknown fractions of the background compositions. 
In the fitted results, the overall simulated background spectrum well describes the measured data with a 9.16-kg NaI(Tl) crystal and shows that the background sources are dominated by surface $^{210}$Pb and internal $^{40}$K in the 2 to 6-keV energy interval, which produce 2.4 counts/day/keV/kg (dru) and 0.5 dru, respectively.
\end{abstract}

\section{Introduction}
\label{intro}
Numerous astronomical observations have led to the conclusion that the majority
of the matter in our universe is invisible, exotic, and
nonrelativistic dark matter \cite{Komatsu:2010fb,Ade:2013zuv}.
However, it is still unknown what the dark matter is. 
Weakly interacting massive particles~(WIMPs) are one of the most attractive
dark matter particle candidates \cite{lee77,jungman96}. 
The lightest supersymmetric particle
(LSP) hypothesized in theories beyond the standard model
of particle physics is a suitable candidate for a dark matter WIMP. There have been numerous experiments that directly search for
WIMPs in our galaxy by looking for nuclear recoils that are produced by WIMP--nucleus scattering ~\cite{gaitskell04,baudis12}. 

To date, no other experiments, except for the DAMA experiment~\cite{bernabei08,bernabei10,bernabei13}, have found an annual modulation signal interpreted as WIMP interactions with a significance of 9.2$\sigma$. However, this finding has spurred a continuing debate since the WIMP--nucleon cross sections inferred from the DAMA modulation are in conflict with limits from other experiments that directly measure the nuclear recoil signals, such as XENON100~\cite{aprile12}, LUX~\cite{agnese14}, and SuperCDMS~\cite{akerib14}. 

The Korea Invisible Mass Search (KIMS) is an experiment that aims at searching for dark matter at an underground laboratory located in Yangyang, South Korea (Y2L). 
We are performing a high-sensitivity search for WIMP interactions in an array of NaI(Tl) crystals in an attempt to reproduce the DAMA/LIBRA's observation of an annual modulation signal~\cite{kims-nai2014,kims-nai2015}. 

There are several groups, such as DM-Ice~\cite{dmice14,dmice16}, ANAIS~\cite{amare14,amare16}, and SABRE~\cite{sabre15}, developing ultra-low-background NaI(Tl) crystals with the goal of reproducing the DAMA/LIBRA results and currently, KIMS and DM-Ice have agreed to operate a single experiment, COSINE, at Y2L using NaI(Tl) crystals and a total mass of 106~kg is being used in the first-stage experiment, COSINE-100. 
As part of this program we have developed ultra-low-background NaI(Tl) crystals and studied their properties in a variety of test setups with the ultimate goal of achieving purity levels that are comparable to or better than those of the DAMA/LIBRA crystals. 
Even though current background levels achieved by the research and development are higher than those of DAMA/LIBRA, 
it is crucial to have a quantitative understanding of the backgrounds.

For further understanding of the backgrounds, 
we have performed Monte Carlo simulations based on Geant4 and compared their results with measured data (see Sect.~\ref{comparison-data}). 
To build concrete background models for a 9.16-kg NaI(Tl) crystal used in one of test setups, 
we studied background simulations of internal radioactive contaminants, such as natural radioisotopes inside NaI(Tl), cosmogenic radionuclei, and surface contaminations in NaI(Tl) crystal (see Sect.~\ref{cosmogenic-isotopes} and~\ref{nai-surface}), and external background sources from the exterior of crystals (see Sect.~\ref{external-background}). 
We quantified their contributions by treating them as floating and/or constrained parameters in the data fitting (see Sect.~\ref{fit}). In addition, our evaluation of background prospects, based on this study, is described in section~\ref{conclusion}.

\section{Experimental setup}

\subsection{Detector shielding and configuration of the test arrangement}
\label{detector-setup}

\begin{figure*}
\begin{center}
\begin{tabular}{cc}
\includegraphics[width=0.5\textwidth]{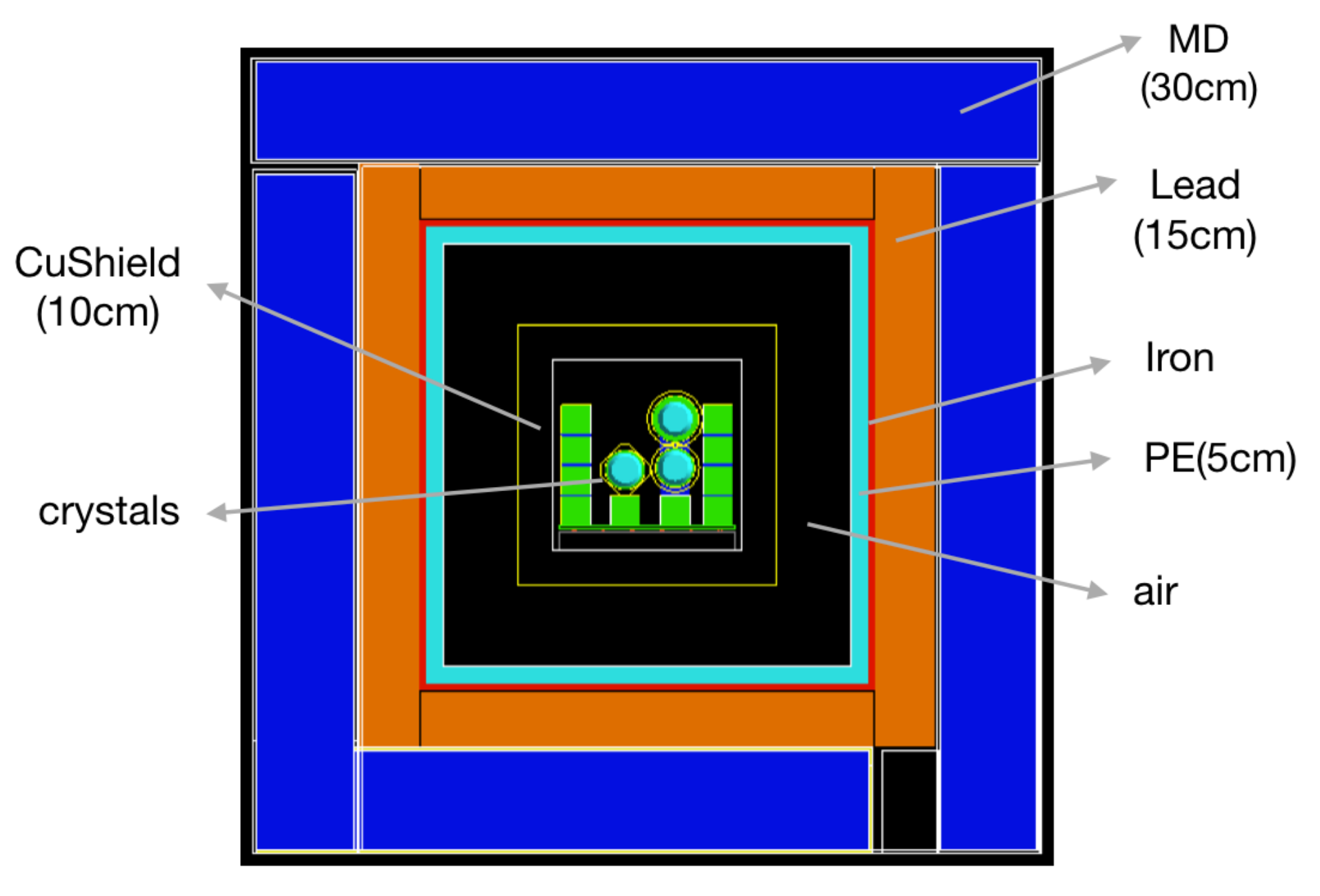} &
\includegraphics[width=0.45\textwidth]{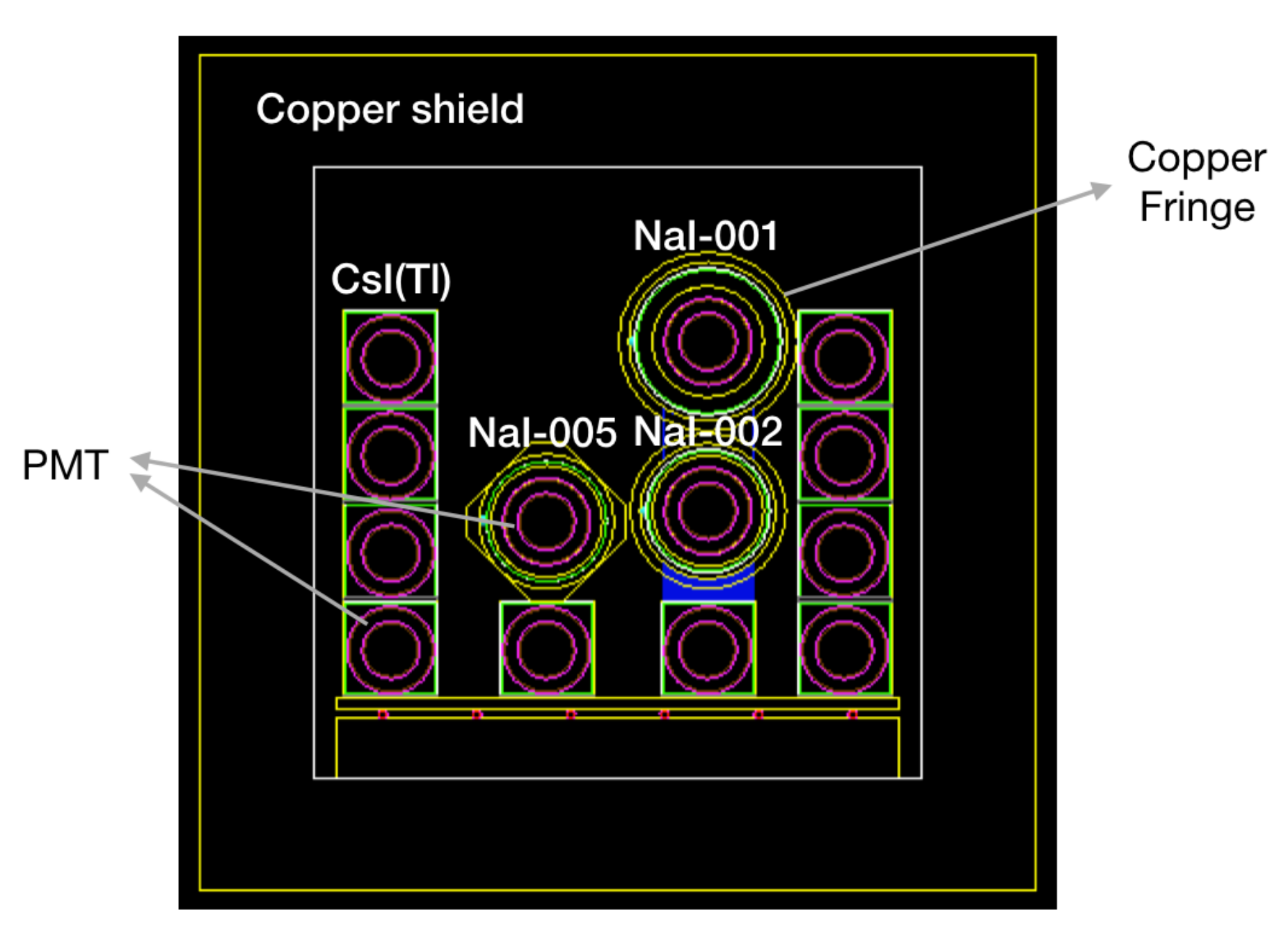} \\
(a) Detector shielding  & (b) Configuration of the test arrangement  \\
\end{tabular}
\caption{Schematic view of detector shielding (a): muon detector~(MD), polyethylene~(PE), and  copper shield~(CuShield). Configuration for three NaI(Tl) crystals with the CsI(Tl) crystal array (b).}
\label{test-setup}
\end{center}
\end{figure*}

We have studied low-background NaI(Tl) crystals with various test setups at Y2L, as reported in references~\cite{kims-nai2014,kims-nai2015}, and as shown in Fig.~\ref{test-setup}(a),
the crystals being tested are enclosed by five different shield layers that were used for the KIMS-CsI experiments~\cite{hslee07,sckim12}.
The outmost layer is a 30-cm-thick muon detector~(MD) filled with mineral oil. The other layers include a sequence of 15-cm-thick lead, an iron sheet, 5-cm-thick polyethylene~(PE), and a 10-cm-thick copper shield.   

One test arrangement that was used for this study is shown in Fig.~\ref{test-setup}(b) with the CsI(Tl) crystal array. 
Three different-sized NaI(Tl) crystals, NaI-001, NaI-002, and NaI-005, are surrounded by ten CsI(Tl) crystals located inside the copper shield. Each end of the crystal was attached to a photomultiplier tube (PMT).
NaI-001, NaI-002, and NaI-005 were produced by Alpa-Spectra (AS) and the first two crystals were grown from their own powders. For NaI-005, it was grown from AS WIMPScint-II grade (AS-WSII) powder that was aimed at reducing $^{210}$Pb contamination in the powder and crystal growing process. As listed in Table~\ref{internalbackgrounds}~\cite{kims-nai2015}, the total $\alpha$ rate of NaI-005 was reduced by more than a factor of three compared to NaI-002.

In this paper, thus, we focused on the background model for the NaI-005 crystal and studied Geant4-based simulations of the background sources internal/external to NaI-005 in this test arrangement. Internal backgrounds of the two crystals, NaI-001 and NaI-002, were simulated for coincidence data of NaI-005 and backgrounds from the PMTs were also simulated for external backgrounds of NaI-005. 

\subsection{Background measurements in the NaI(Tl) crystal test setup}

\begin{table*}
\begin{center}
\caption{Light yield and measured background rates from internal radioactive contaminants in the NaI(Tl) crystal~\cite{kims-nai2015}.
}
\label{internalbackgrounds}
\begin{tabular}{ccccccc} \hline
 Crystal  &$^{nat}$K($^{40}$K)      & $^{238}$U & $^{232}$Th  & $\alpha$ Rate & Light yield        \\
  (unit)   & (ppb)        & (ppt)            & (ppt)           &  (mBq/kg)                      & (PE/keV)           \\
  \hline 
  NaI-001  & $40.4\pm2.9$ &  $<0.02$         & $<3.19$         & $3.29\pm0.01$   &  $15.6\pm1.4$    \\
  NaI-002  & $48.1\pm2.3$ &  $<0.12$         & $0.5\pm0.3$     & $1.77\pm0.01$   &  $15.5\pm1.4$   \\
  NaI-005  & $40.1\pm4.2$ &  $<0.04$         & $0.19\pm0.01$  & $0.48\pm0.01$    & $12.1\pm1.1$     \\ \hline
\end{tabular}
\end{center}
\end{table*}

\begin{table*}
\begin{center}
\caption{Specifications for PMTs used in this study~\cite{kims-nai2014}. (a) The radioactivities were measured with a HPGe detector at Y2L. (b) SEL means ``selected for high quantum efficiency''.
}
\label{photomultipliertubes}
\begin{tabular}{c|c|c|c||c}
\hline
\multicolumn{2}{c|}{PMT}  & $\rm R12669SEL^{b}$   & $\rm R11065SEL^{b}$  & 9269QA  \\ \hline 
\multicolumn{2}{c|}{Photocathode}  & SBA     &  Bialkali  & RbCs\\
\multicolumn{2}{c|}{Window} & Borosilicate     & Quartz  & \\ 
\multicolumn{2}{c|}{Body} &  Borosilicate    & Kovar  & \\ 
\multicolumn{2}{c|}{Stem} &  Glass    & Glass  & \\ \hline
$\rm Radioactivity^{a}$     & U($^{214}$Bi ) &  25 $\pm$ 5    &  60
$\pm$ 5  & 78.2 $\pm$  4.2 \\ 
(mBq/PMT)       & Th($^{228}$Ac) &  12 $\pm$ 5    & 0.5
$\pm$ 0.2 & 25.5 $\pm$ 4.4 \\ 
                       & K($^{40}$K) &  58 $\pm$ 5    & 19
                       $\pm$ 2  & 504 $\pm$  72 \\  
\hline
\end{tabular}
\end{center}
\end{table*} 

NaI-002 and NaI-005 crystals have an identical cylindrical shape with a diameter of 4.2 inches, a length of 11 inches, and a mass of 9.16~kg. 
The light yield and the measured background rates from internal radioactive contaminants in the NaI(Tl) crystals are listed in Table~\ref{internalbackgrounds}. 
The details of the three crystals are discussed in reference~\cite{kims-nai2015}.

Three different types of PMTs were used in the test arrangement: a metal-packed R11065SEL, a glass-packed R12669SEL, both manufactured by Hamamatsu Photonics, and 9269QA of Electron Tubes, Ltd.. \\
R12669SEL PMTs were coupled with NaI-002 and NaI-005; R11065SEL and 9269QA were coupled with NaI-001 and CsI(Tl) crystals.
The radioactivity levels of the PMTs were measured underground with a high-purity Ge (HPGe) detector and their measurements are listed in Table~\ref{photomultipliertubes}~\cite{kims-nai2014}. 

We used the measured activities inside the crystals and from the PMTs for the simulation study of the NaI-005 backgrounds.
In addition to natural radioisotopes measured, there are also backgrounds from cosmic excitation that are continuously decreasing as a function of time. We considered I, Te, and $^{22}$Na isotopes from cosmic radiation as background sources and they are included in the simulations and their contribution described later.

\section{Background simulations}

\subsection{Method of simulation}

\begin{table*}
\begin{center}
\caption{Isotopes grouped by half-life: five groups for  $^{238}$U (a) and three groups for  $^{232}$Th (b).}
\label{group-u238-th232}
\begin{minipage}{0.47\textwidth}
\begin{tabular}{c|ccc}
 \multicolumn{4}{c}{(a)} \\ \hline
 Group & \multicolumn{2}{c}{Decay chain} & Half-life \\ 
 & & & \\ \hline 
 & $^{238}$U & $^{234}$Th & 4.47$\times10^{9}$ yr \\
1 & $^{234}$Th & $^{234}$Pa & 24.1 days \\
 & $^{234}$Pa & $^{234}$U & 6.70 h \\ \hline
2 & $^{234}$U & $^{230}$Th & 2.46$\times10^{5}$  yr \\ \hline
3 & $^{230}$Th & $^{226}$Ra & 7.54$\times10^{4}$  yr \\ \hline
 & $^{226}$Ra & $^{222}$Rn & 1.60$\times10^{3}$  yr \\ 
 & $^{222}$Rn & $^{218}$Po & 3.82 days \\
4 & $^{218}$Po & $^{214}$Pb &  3.10 min \\
 & $^{214}$Pb & $^{214}$Bi & 26.8 min \\
 & $^{214}$Bi & $^{214}$Po & 19.9 min \\
 & $^{214}$Po & $^{210}$Pb & 1.64$\times10^{-6}$  s \\ \hline
 & $^{210}$Pb & $^{210}$Bi & 22.2 yr \\
5 & $^{210}$Bi & $^{210}$Po & 5.01 days \\
 & $^{210}$Po & $^{206}$Pb & 138 days \\  \hline 
\end{tabular}
\centering
\end{minipage}
\hfill
\begin{minipage}{0.47\textwidth}
\begin{tabular}{c|ccc}
 \multicolumn{4}{c}{(b)} \\ \hline
 Group & \multicolumn{2}{c}{Decay chain} & Half-life \\ 
 &  & &  \\ \hline 
6 & $^{232}$Th & $^{228}$Ra & 1.40$\times10^{10}$ yr \\ \hline
7 & $^{228}$Ra & $^{228}$Ac & 5.75 yr \\
 & $^{228}$Ac & $^{228}$Th & 6.15 hr \\ \hline
 & $^{228}$Th & $^{224}$Ra & 1.91 yr \\ 
 & $^{224}$Ra & $^{220}$Rn & 3.63 days \\
 & $^{220}$Rn & $^{216}$Po & 55.6 s \\ 
 & $^{216}$Po & $^{212}$Pb & 0.145 s \\
8 & $^{212}$Pb & $^{212}$Bi &  10.6 hr \\
 & $^{212}$Bi & $^{212}$Po & 60.6 min \\
 & $^{212}$Po & $^{208}$Pb & 2.99$\times10^{-7}$ s \\
 & $^{212}$Bi & $^{208}$Tl & 60.6 min \\
 & $^{208}$Tl & $^{208}$Pb & 3.05 min \\ \hline 
\end{tabular}
\end{minipage}
\end{center}
\end{table*}

To understand the measured background of NaI-005 in the test arrangement described in section~\ref{detector-setup},
we have performed Monte Carlo simulations with the Geant4 toolkit~\cite{geant4}, version of 4.9.6.p02, including physics list classes of G4EmLivermorePhysics for low energy electromagnetic process and G4RadioactiveDecay for radioactive decay process.

In the simulation framework, radioactive sources such as full decay chains of $^{238}$U and $^{232}$Th were simulated, assuming that decay chains of $^{238}$U and $^{232}$Th are each in equilibrium, thus all related activities within the chains are simply equal to the $^{238}$U and $^{232}$Th activities multiplied by the branching ratios for decay of the daughter isotopes. However, it needs to specify all long-lived parts of $^{238}$U and $^{232}$Th daughters to consider broken decay chains when they are fitted to the measured data to quantify their unknown background fractions
and, thus, we grouped daughter isotopes from the full decay chains of $^{238}$U and $^{232}$Th according to their half-lives.  
We used five groups for $^{238}$U and three groups for $^{232}$Th, as listed in Table~\ref{group-u238-th232}, in the data fitting. 
$^{40}$K is treated as its own, additional group.

Each simulated event includes all energy deposited in the crystals within an event window of 10 $\mu$s from the time a decay is generated, to account for the conditions in the data acquisition system (DAQ) of the experimental setup~\cite{kims-nai2014}.  
Sometimes decays with relatively short half-life such as $^{212}$Po decay (with a half-life of 300~ns) and the following decays will appear in the same event, called pileup events, and they were treated within one event in simulated event. 
 
The simulated spectrum was smeared by the energy resolution, as a function of energy, that was measured during the calibration. 
Calibration points were measured using $\gamma$--ray sources, $^{241}$Am and $^{60}$Co; they contribute peaks at 59.5~keV($^{241}$Am), 1170~keV, and 1332~keV.
Measured background spectra from several radioactivities were used for additional calibrations; there are peaks at 3.2~keV and 1460~keV from $^{40}$K, 30.8~keV from $^{126}$I, 
67.2~keV from $^{125}$I, and 609.3~keV and 1120.3 keV from $^{214}$Bi, respectively. 
We also used the peak at 0.8~keV from $^{22}$Na.

\subsection{Comparison of simulated background spectra with measured data}
\label{comparison-data}

\begin{figure}[!htb]
\begin{center}
\includegraphics[width=0.5\textwidth]{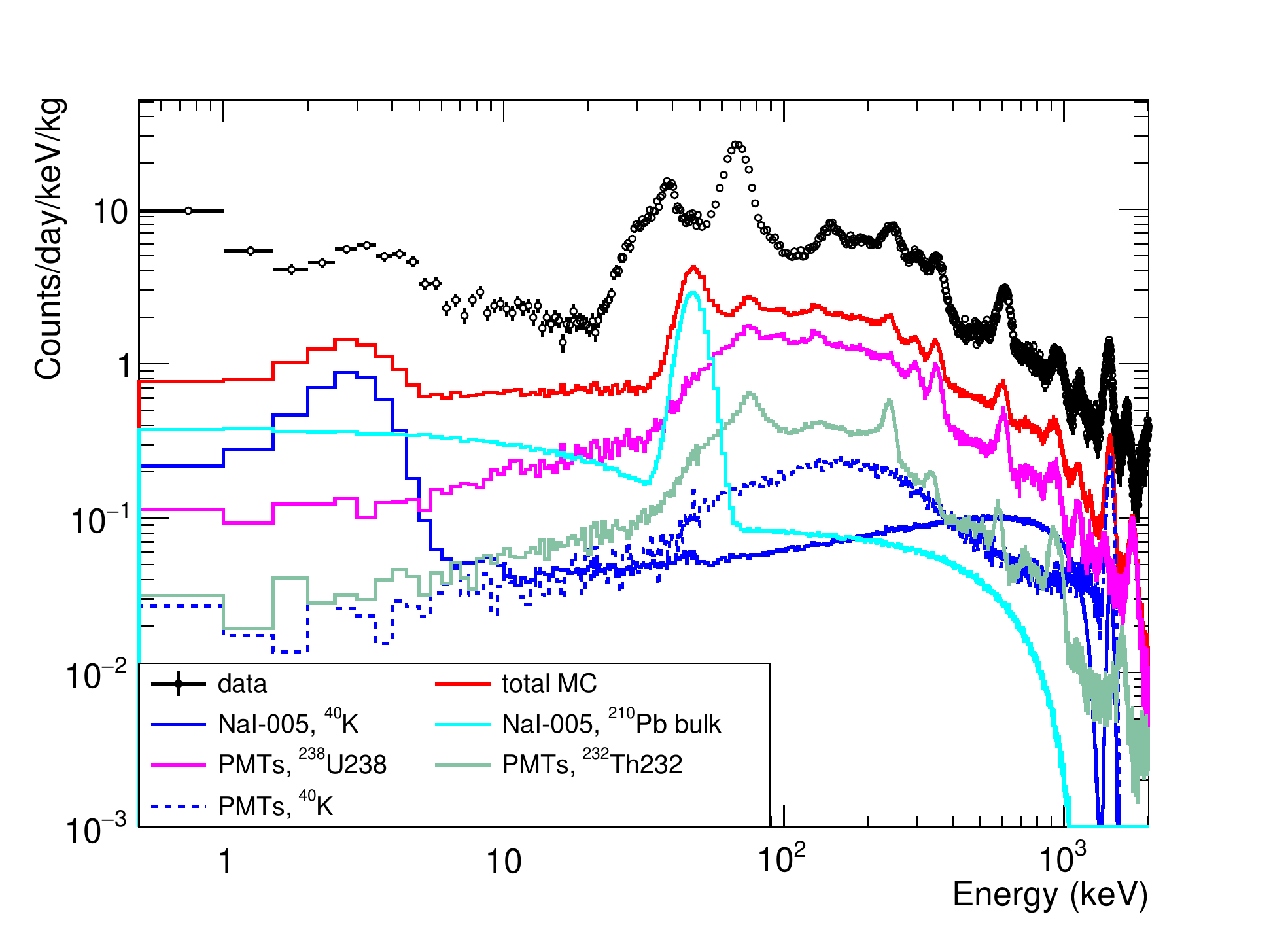}
\caption{
Comparison of measured background spectra to spectra generated by Monte Carlo simulation for single-hit events of NaI-005 
}
\label{comparison}
\end{center}
\end{figure}

To build the concrete background model for NaI-005
it needs to compare the measured background level in NaI-005 with the simulated background spectra that are each composed of different background compositions.
Initially, we studied background simulations of natural radioisotopes in NaI(Tl) crystals and PMTs and compared the total of the simulations with the data.
In this comparison, we found there remains significant backgrounds to be modeled and, thus, we considered additional background components to improve the model. The details of the simulation studies are described in subsections. 

For the initial comparison,
we simulated daughter isotopes from full decay chains of $^{238}$U, $^{232}$Th, and  $^{40}$K located inside NaI(Tl) crystals and 26 PMTs assuming a chain equilibrium.
By using the measured activities of radioactive sources listed in Tables~\ref{internalbackgrounds} and \ref{photomultipliertubes}, event rates were normalized to the units of counts/day/keV/kg (dru). 
Fig.~\ref{comparison} shows the normalized background energy distributions in NaI-005 for single-hit events in which there is an energy deposit in NaI-005 only.
The total of the simulations (solid red line) is compared with the measurement (open black circles).
In the simulations, the background sources for energies below 10~keV are dominated by $^{40}$K (solid blue line) and $^{210}$Pb (solid cyan line) internal to the NaI(Tl) crystal, and for high energies above $\sim$100~keV external backgrounds from PMTs are dominant. The details are itemized in the following.

However, as shown in Fig.~\ref{comparison}, the overall measured background level is higher than that of background simulations and, thus, there are significant remaining backgrounds.
For the contributions of the background sources not included in Fig.~\ref{comparison}, we studied background simulations of the external sources, cosmogenic radionuclei, and surface contaminations of the NaI(Tl) crystal.
To quantify the contributions of all background sources in the simulation they are fitted to the measured data by floating and/or constrained the unknown fractions of the background spectra as fit parameters.
In the fit, daughter isotopes from full decay chains of $^{238}$U, $^{232}$Th, and  $^{40}$K are grouped into 9 background spectra: 5 background spectra for $^{238}$U, 3 background spectra for $^{232}$Th, and a spectrum for $^{40}$K.
The fitted results for the contributions of those background sources to the total background level are described in sections~\ref{external-background}, \ref{cosmogenic-isotopes}, and \ref{nai-surface}, quantitatively. All the fit results of activities of background sources and background events in the 2 to 6~keV energy region are listed in Tables~\ref{fit-result} and~\ref{fit-result-lowenergytab} in section~\ref{fit}.

\begin{itemize}
  \item[$\bullet$] Internal backgrounds of NaI(Tl) crystals
  
  To normalize the backgrounds from internal radioactive contaminants, we assumed a chain equilibrium. Therefore, all related activities within the chains are equal to $^{238}$U, $^{232}$Th, and $^{40}$K activities, in Table~\ref{internalbackgrounds}, multiplied by the branching ratios for decay of the daughter isotopes. We also added the background simulation of internal $^{210}$Pb by considering the measured $\alpha$ rate. The resultant background contributions, except for those from $^{40}$K and $^{210}$Pb, were negligible ($<10^{-2}$ dru). \\
    
  \item[$\bullet$] External backgrounds from PMTs 
  
  We used 26 3-inch PMTs of three different PMT types in the test arrangement. We thus considered measured activities in terms of different PMT types, as listed in Table~\ref{photomultipliertubes}. 
The normalized background contributions from the PMTs are represented in three radioactive sources
: $^{238}$U (solid magenta line), $^{232}$Th (solid khaki line), and $^{40}$K (dotted blue line). 
\end{itemize}

\subsubsection{Effects of external background sources}
\label{external-background}

\begin{figure*}
\begin{center}
\begin{tabular}{cc}
\includegraphics[width=0.5\textwidth]{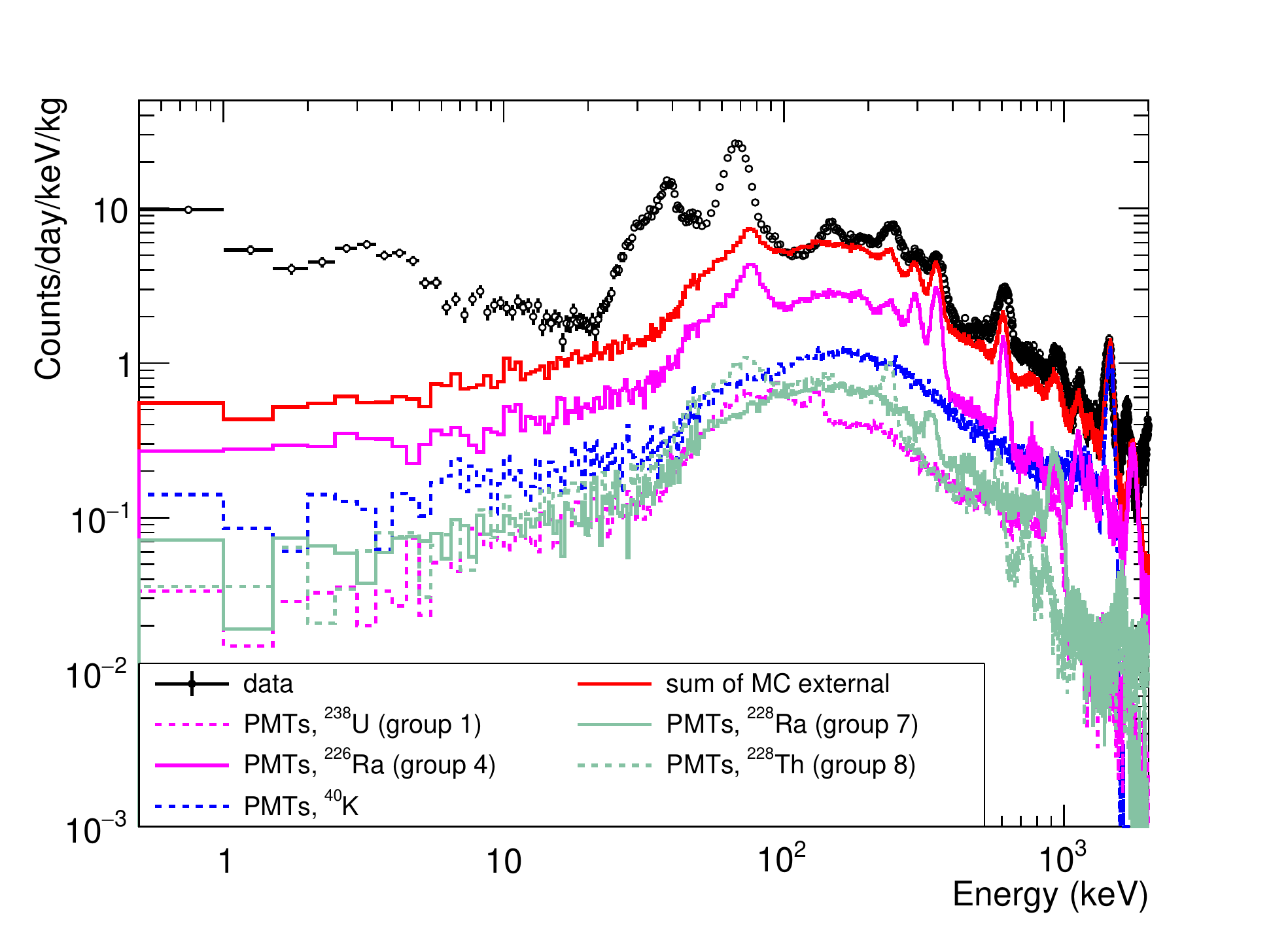} &
\includegraphics[width=0.5\textwidth]{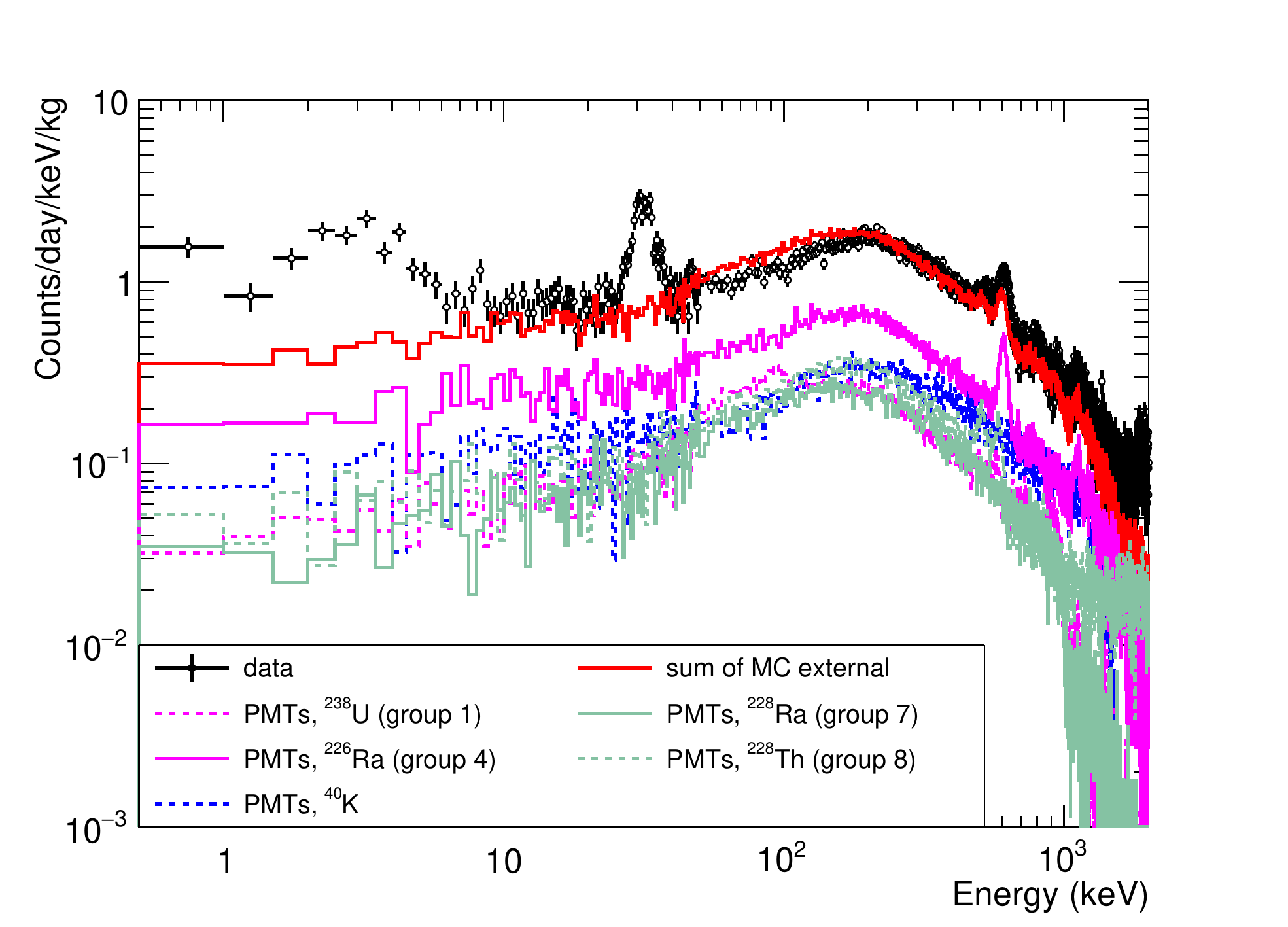} \\
(a) Single-hit events  & (b) Multiple-hit events \\
\end{tabular}
\caption[]{Simulated background spectra of $^{238}$U, $^{232}$Th, and $^{40}$K from external sources inside the copper shield.}
\label{external-bkg}
\end{center}
\end{figure*}

The majority of the external $\gamma$ background comes from the radioactive decay of isotopes in the surrounding rocks.
To block such an environmental background we installed several shielding layers, as described in section~\ref{detector-setup}, and measured the background reduction with an ultra-low background 100\% HPGe detector at Y2L. By using the full shielding structure and $N_{2}$ gas flowing into the inside of the copper shield to avoid backgrounds from $^{222}$Rn in the air at Y2L, we could reduce the environmental background by a factor of 10,000 (measured to be $1.20\pm 0.49$ pCi/L~\cite{kims-radon2011}), thus ensuring that those contributions would be negligible.

However, there exist some background from radioactive sources in detector components inside the shielding.
Typically, PMTs and materials---such as the copper shield, the copper structure to which the PMT is affixed to the crystal, the PMT base, connectors, and bunches of cables---would contain radioactive sources that can contribute to the background as an external background source inside the shielding.
For PMTs, even though we measured the radioactivity levels, background contributions will be different for the main components in the PMTs, such as the PMT window, body, and stem, where radioactive sources were generated. 

Although it is difficult to reproduce accurately background contributions from external sources without knowing well about the external background contamination, it is possible to estimate the effects of such radioactive sources, external to the NaI(Tl) crystal, by considering them as parameters in the fit. 
To take it into account we simulated the background spectra in such a way that radioisotopes contained in $^{238}$U, $^{232}$Th, and $^{40}$K were generated randomly in the whole PMT body including the PMT window and stem, and we grouped the results into nine background spectra, instead of performing each of the simulations of all the external sources besides PMTs.
It is because that the background energy distribution in NaI-005 is very similar to that from PMTs when we simulated external background sources from the inside space of the copper shield.
We also separately considered background contributions from the two PMTs attached to NaI-005, apart from the other PMTs attached to NaI-001, NaI-002, and CsI(Tl) crystals, because the external background contributions to NaI-005 will be different for the distance from the external sources to the crystal. 
Consequently, we have two sets of nine background spectra for 
testing the effects of the external background sources.
They were fitted to the measured data, as floating and/or constrained parameters, to estimate their fractions contributed to the total background and, in the fit, we used single-hit events and multiple-hit events, the latter of which have energy deposits in two or more crystals, simultaneously.
We estimated the external background contributions by assuming that the fitted results for PMTs simulations include the backgrounds from all the external sources.
As shown in Fig.~\ref{external-bkg}, the fitted results well reproduce the measured data for both single-hit events and multiple-hit events for energies above $\sim$100~keV. 
The black circles represent the data and the solid red line represents the sum of simulated background spectra with the fitted fractions of groups 1 (dotted magenta line), 4 (solid magenta line), 7 (solid khaki line), and 8 (dotted khaki line) and $^{40}$K (dotted blue line); the biggest contribution is from $^{226}$Ra (group 4). Those external backgrounds are expected to be vetoed by an active veto detector in the new detector design.

\subsubsection{Backgrounds from cosmogenic isotopes}
\label{cosmogenic-isotopes}

\begin{figure}[!htb]
\begin{center}
\includegraphics[width=0.5\textwidth]{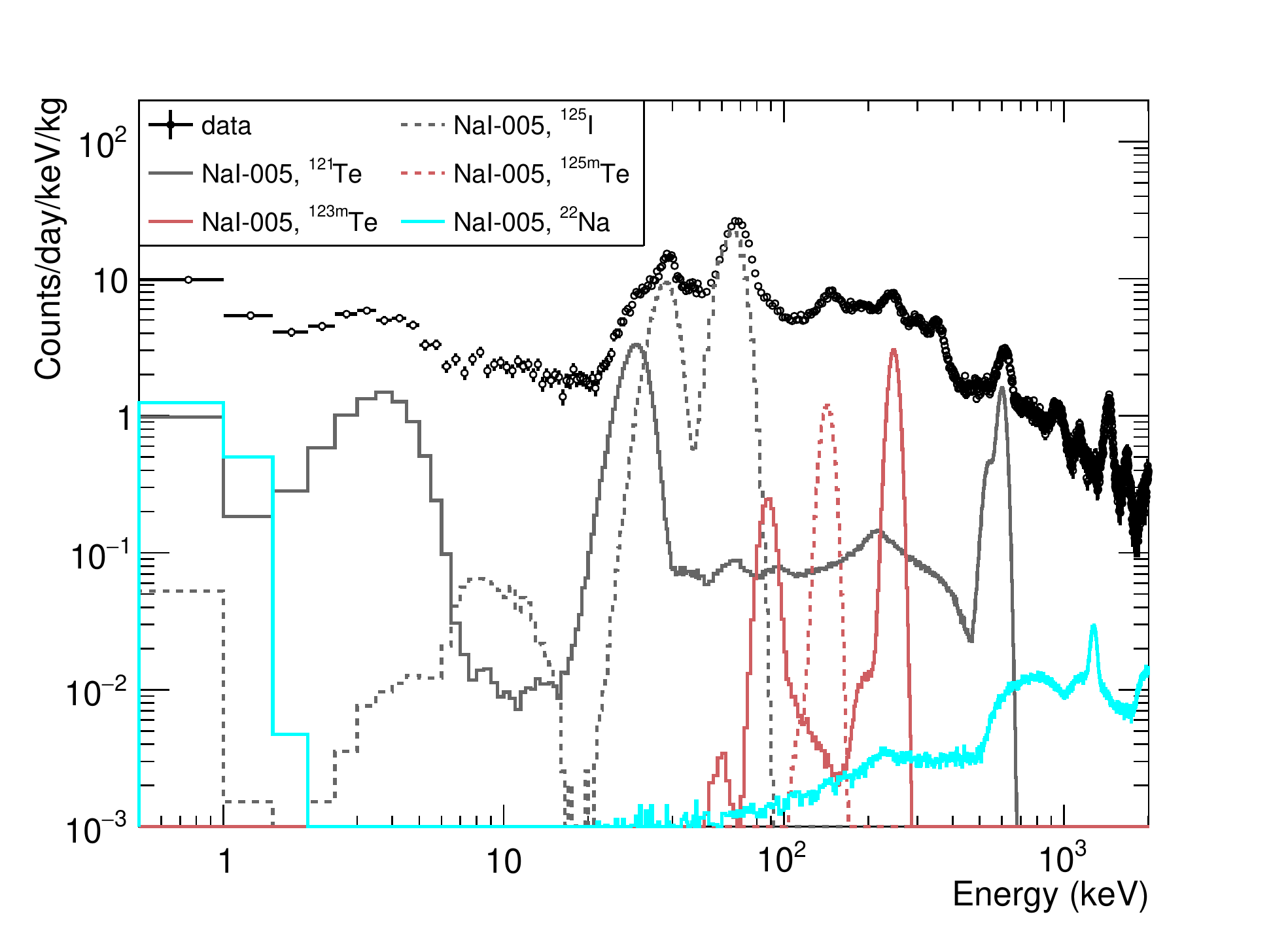}
\caption{Cosmogenic backgrounds for the NaI(Tl) crystal with the CsI(Tl) crystal array.}
\label{cosmogenic}
\end{center}
\end{figure}

We installed NaI-005 in the test setup in December 2014, soon after that crystal was delivered to Y2L, and we recorded data for about 
a month. Therefore, some backgrounds are expected from cosmogenic activations. 
We checked backgrounds over the specified time interval and observed a clear reduction of the peak at 30.8~keV from $^{126}$I within the first 10 days.  
We thus did not use the first 16 days data to exclude the background from $^{126}$I in this study.
We simulated backgrounds from cosmogenic isotopes, which are expected to be produced by cosmic ray exposure~\cite{walter-thesis}. The simulated background spectra are used in the data fitting, by floating their unknown fractions, and the fitted results are shown in Fig.~\ref{cosmogenic}. The dominant isotopes are itemized in detail in the following.

\begin{itemize}
  \item[$\bullet$] $^{125}$I (dotted gray line)
  
 $^{125}$I has a half-life of 59.43 days and it decays to $^{125}$Te by electron capture from shell K and upper shells, emitting 35.5 and 
67.2-keV $\gamma$ rays and/or internal conversion electrons, respectively, producing the two big energy peaks shown by the dotted gray line in Fig.~\ref{cosmogenic}. \\ 
 
  \item[$\bullet$] $^{121}$Te (solid gray line)
  
  The half-lives of $^{121}$Te and $^{126}$I are 19.17 and 12.93 days, respectively, and a clear reduction of the peak from $^{126}$I within the first 10 days was observed. 
  However, there is still peak at 573~keV identified with $\gamma$ rays emitted by electron capture in the decay of $^{121}$Te  
  that is produced as a result of the decay of the longer-lived cosmogenic isotope $^{121m}$Te (half-life = 164.2 days), which is reported in reference to ANAIS~\cite{amare15}. \\   
  
  \item[$\bullet$] $^{125m}$Te (dotted brown line) and $^{123m}$Te (solid brown line)
  
  $^{125m}$Te and $^{123m}$Te are long-lived metastable states and their half-lives are 57.4 and 119.2 days, respectively. They contribute peaks at 145 and 248~keV. \\
  
  \item[$\bullet$] $^{22}$Na (solid cyan line)

  $^{22}$Na can be produced through the $(n, 2n)$ reaction on $^{23}$Na by energetic cosmic neutrons at sea level. 
It decays via positron emission (90\%) and electron capture (10\%), followed by 1270-keV $\gamma$-ray emission with a mean lifetime of 3.8 yr. 
The electron capture decay produces $\sim$0.8-keV X-rays. Therefore, $\sim$10\% of the $^{22}$Na decay will produce 0.8-keV X-rays and 1270-keV $\gamma$ rays simultaneously. Meanwhile, the positron will be converted to two 511-keV annihilation $\gamma$ rays. If one of the two 511 and 1270~keV $\gamma$ rays escapes from a crystal, the energy deposited in the crystal will be 650--1000~keV.  We looked for a coincident event that deposits 1270~keV of energy in NaI-002, resulting in a $\gamma$-ray hit in the 650 to 1000-keV energy interval in NaI-005. It is measured to be $0.8\pm 0.3$~mBq/kg by considering the detection efficiency obtained from a Monte Carlo simulation and is used as a constrained parameter for the background fraction of $^{22}$Na in the fit. 
\end{itemize}

\subsubsection{Background from NaI(Tl) surface}
\label{nai-surface}

\begin{figure}[!htb]
\begin{center}
\includegraphics[width=0.45\textwidth]{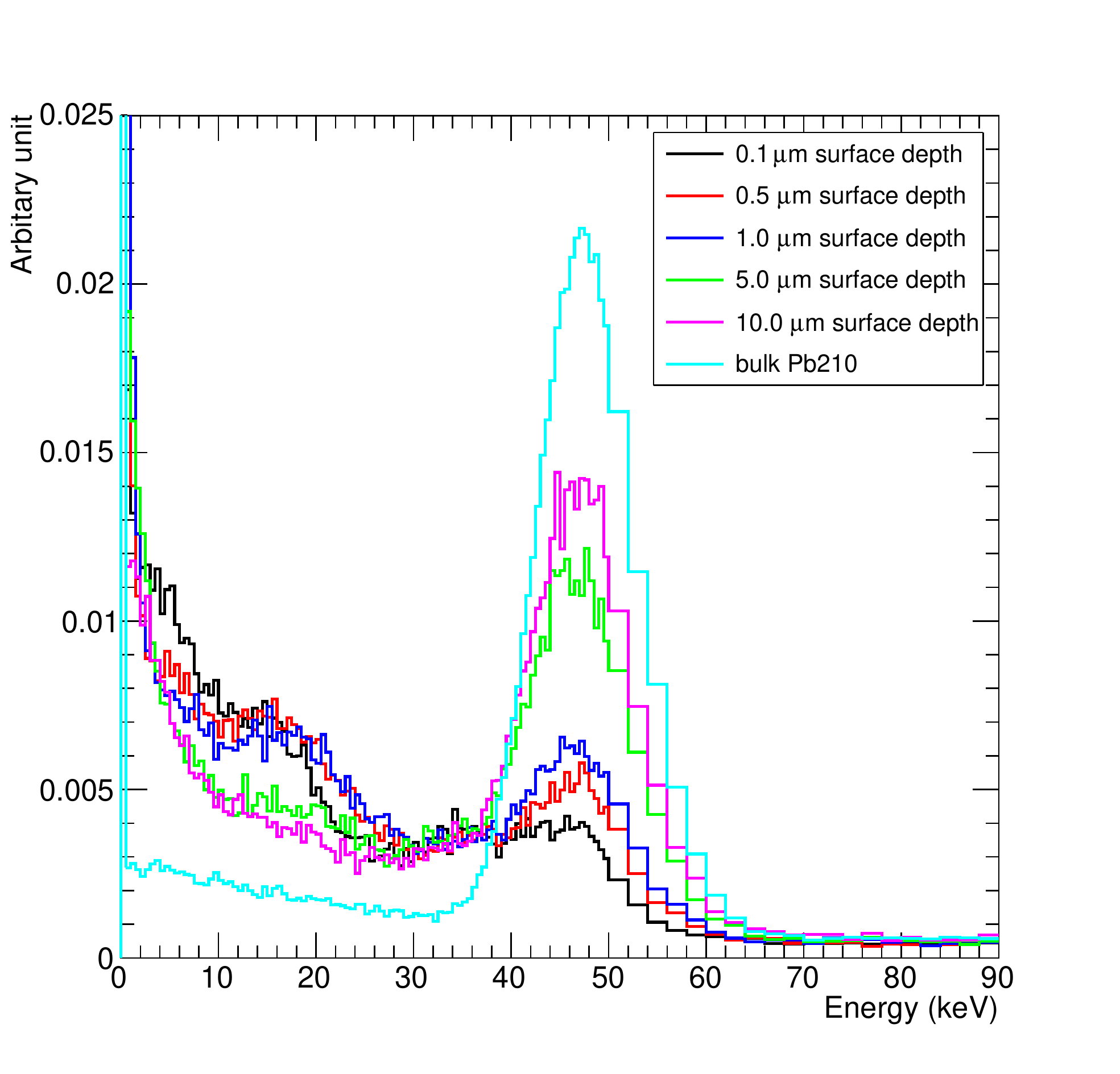}
\caption{Comparison of background spectra simulated at various surface depths.}
\label{surfacedepth}
\end{center}
\end{figure}

\begin{figure}[!htb]
\begin{center}
\includegraphics[width=0.5\textwidth]{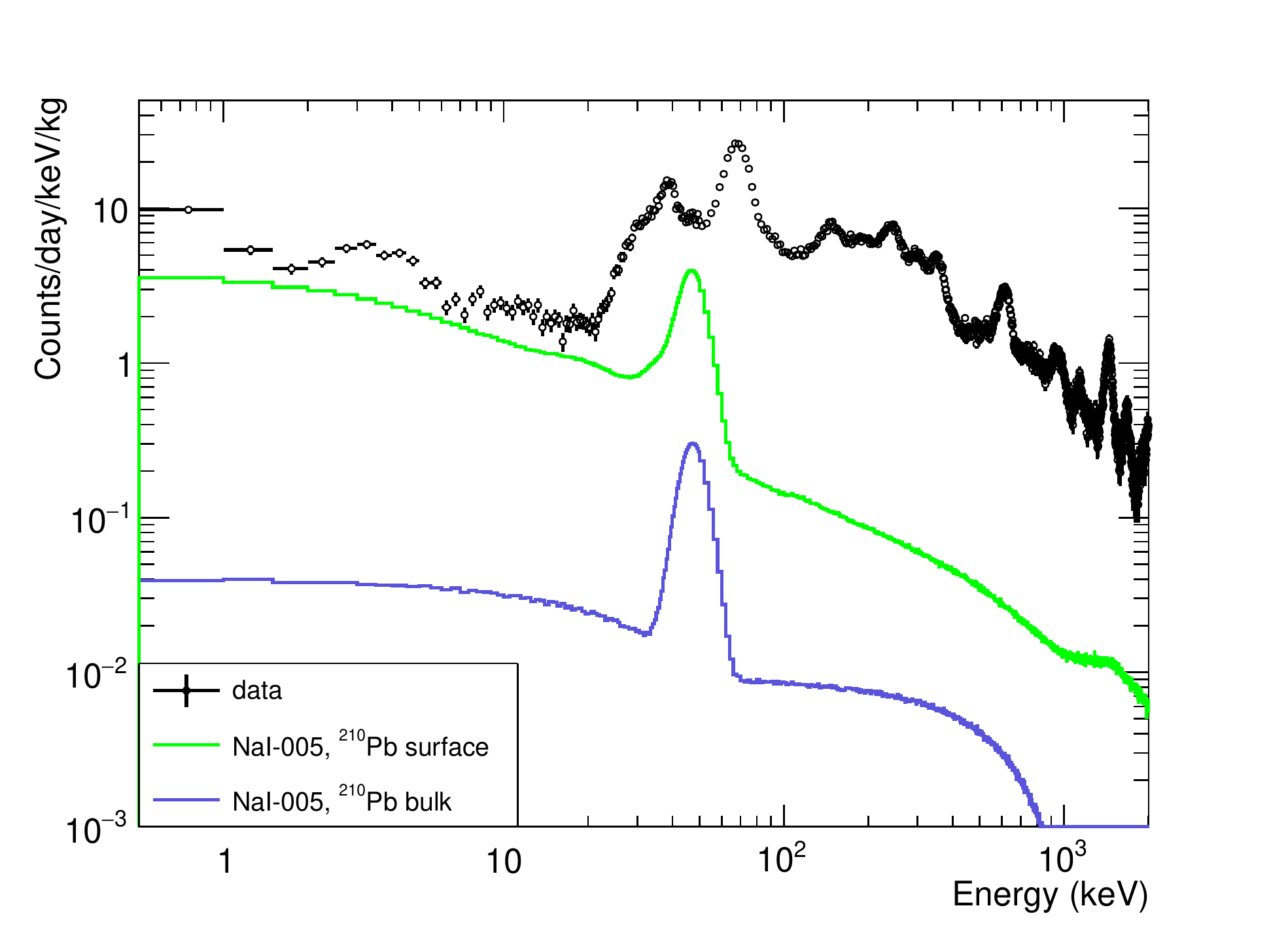}
\caption{Simulated background spectra from bulk $^{210}$Pb (solid blue line) and a $^{210}$Pb surface (solid green line).}
\label{surface}
\end{center}
\end{figure}

The levels of $^{238}$U and $^{232}$Th contamination measured in NaI-005 are too low to account for the total observed $\alpha$-particle rate, which suggests that they are due to decays of $^{210}$Po nuclei that originate from $^{222}$Rn contamination that occurred sometime during the powder and/or crystal processing stages. 
This is confirmed by the observation of a 46-keV $\gamma$ peak that is characteristic of $^{210}$Pb.
There is also the possibility of surface contamination, which is expected to affect the low-energy spectrum in a different way from bulk $^{210}$Pb.  

To clarify the effect of  surface $^{210}$Pb contamination we simulated the background spectra at various surface depths from 0.1 to 10~$\mu$m by generating $^{210}$Pb randomly within the depth. Fig.~\ref{surfacedepth} shows a comparison of background spectra simulated in different surface depths. 
The low-energy background looks different from that due to bulk $^{210}$Pb (cyan color), and the heights of the $\sim$50-keV peak are also different from that due to bulk $^{210}$Pb. 

In the simulation, we added a background contribution from surface $^{210}$Pb generated randomly within a surface depth of 10~$\mu$m.
Background fractions of bulk and surface $^{210}$Pb are obtained in such a way that they are treated as floating parameters in the fit to the measured data. 
And they are found to be 0.05$\pm$0.76 and 0.81$\pm$0.25 mBq/kg, respectively, as shown in Fig.~\ref{surface}. The total amount of 0.86$\pm$0.80 mBq/kg is comparable to the total observed $\alpha$ activity from decays of $^{210}$Po from crystals grown by AS-WSII powder ($0.85\pm 0.04$ mBq/kg).
 
\subsubsection{Comparison of fitted Monte Carlo fractions with data}
\label{fit}

\begin{figure*}
\begin{center}
\begin{tabular}{cc}
\includegraphics[width=0.5\textwidth]{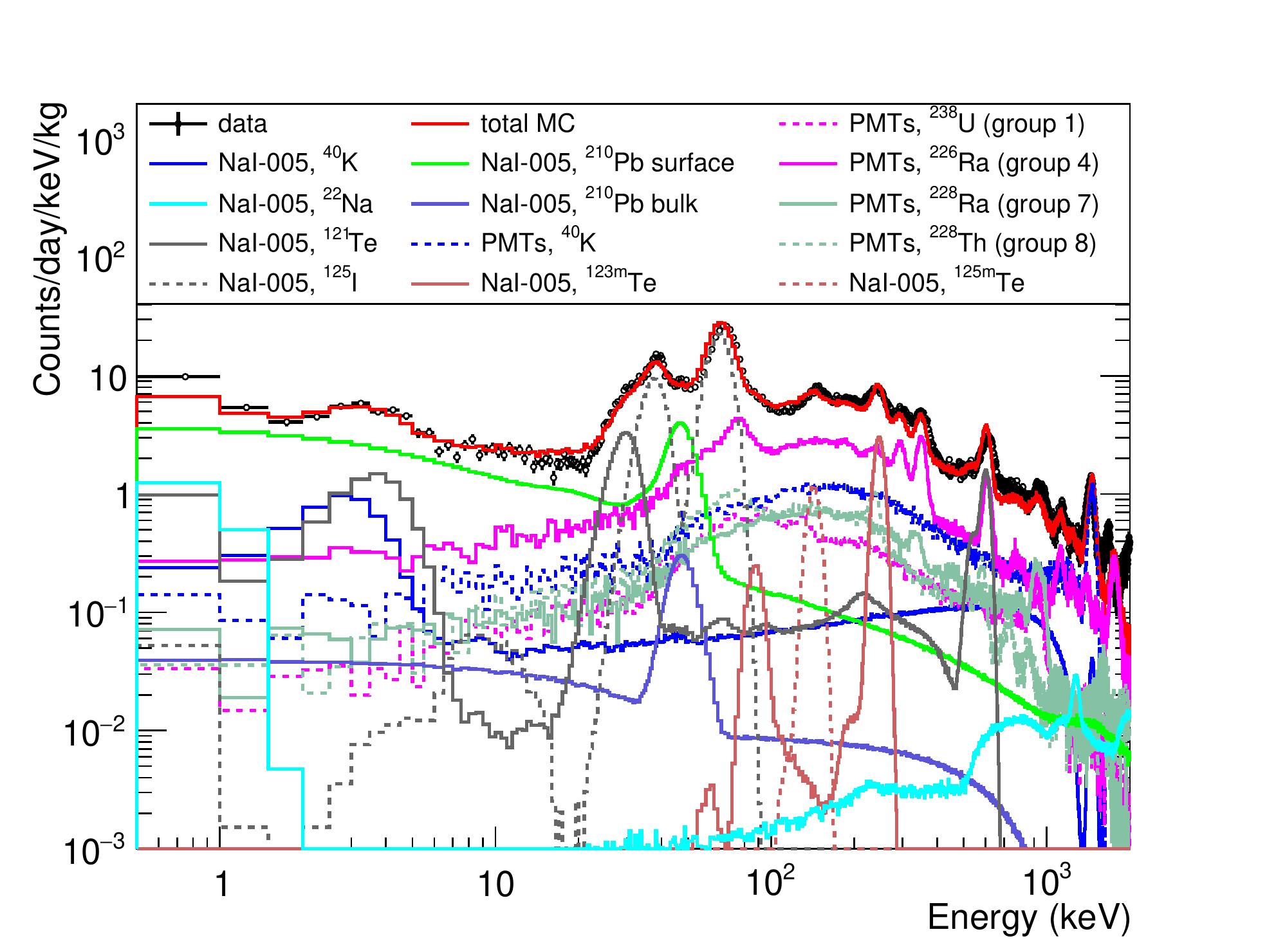} &
\includegraphics[width=0.5\textwidth]{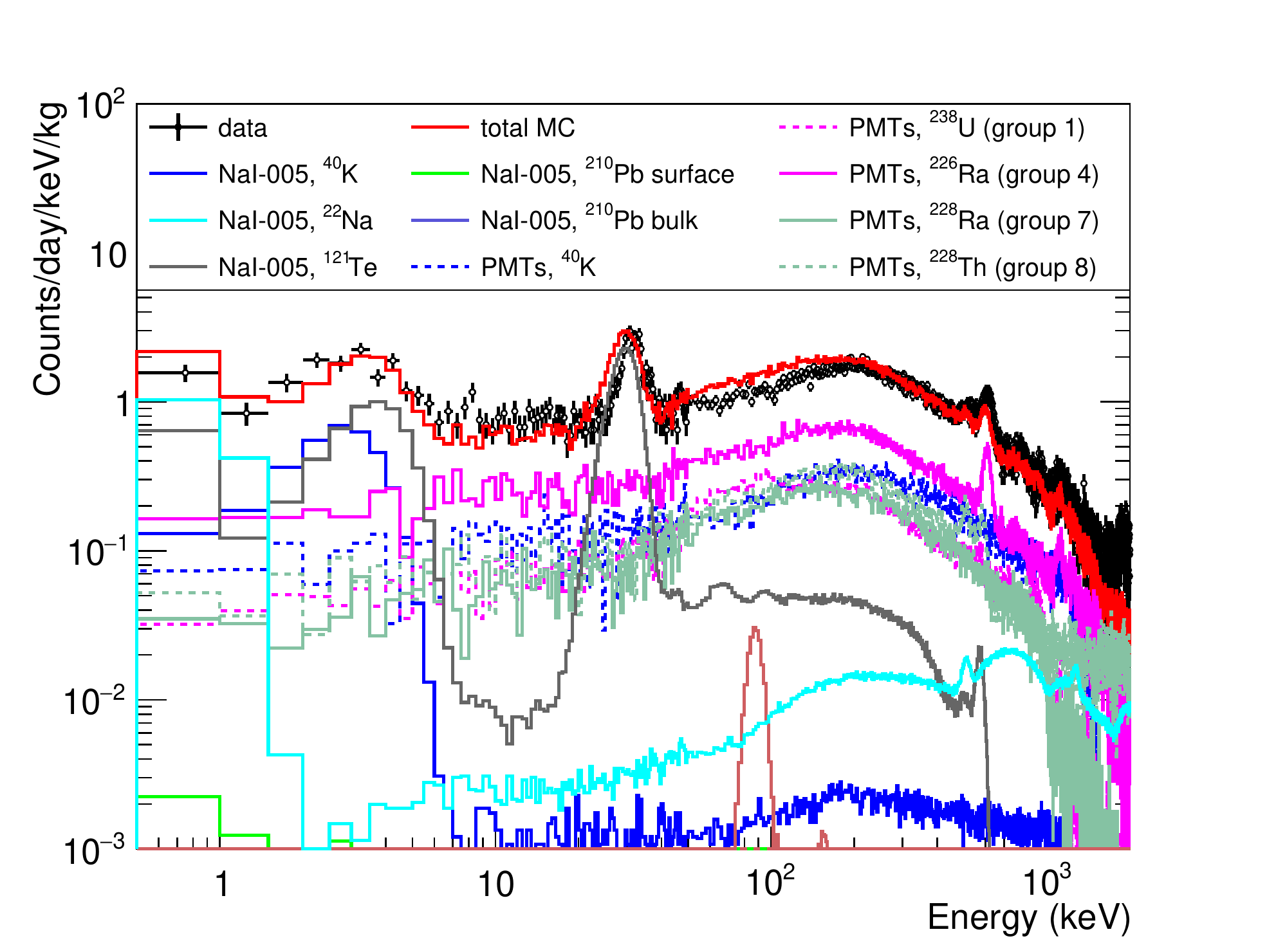} \\
(a) & (b) \\ 
\end{tabular}
\caption[]{Measured background spectrum (open black circles) fitted with all of simulated background spectra for single-hit events (a) and multiple-hit events (b).
}
\label{result}
\end{center}
\end{figure*}

\begin{table*}
\begin{center}
\caption{Summary of the fitted radioactive contaminants in NaI-005 (a) and PMTs (b).}
\label{fit-result}
\begin{tabular}{ccccc}
\multicolumn{5}{c}{(a)} \\ \hline
 & & & \multicolumn{2}{c}{Activities} \\
 & & & \multicolumn{2}{c}{[mBq/kg]} \\  \cline{4-5}
 Background sources & Group & Isotopes & Measured & Fitted \\ \hline
 & 1 & $^{238}$U & $<5.0\times10^{-4}$ & $(5.50\pm0.87)\times10^{-4}$ \\
 & 2 & $^{234}$U & & $(5.50\pm0.67)\times10^{-4}$ \\ 
 & 3 & $^{230}$Th & & $(5.50\pm0.41)\times10^{-4}$ \\ 
 & 4 & $^{226}$Ra & & $(5.50\pm1.01)\times10^{-4}$ \\
 Internal & 5 & $^{210}$Pb & & $(4.51\pm0.40)\times10^{-4}$ \\
 & 6 & $^{232}$Th & $(7.79\pm0.41)\times10^{-4}$ & $(8.80\pm0.87)\times10^{-4}$ \\ 
 & 7 & $^{228}$Ra & & $(7.20\pm0.72)\times10^{-4}$ \\
 & 8 & $^{228}$Th & & $(8.80\pm1.10)\times10^{-4}$ \\
 & 9 & $^{40}$K & $1.20\pm0.13$ & $1.32\pm0.21$ \\ \hline
 Bulk &  & $^{210}$Pb & & $0.05\pm0.76$ \\
 Surface &  & $^{210}$Pb & & $0.81\pm0.25$  \\ \hline
 &  & $^{125}$I & & $5.01\pm0.51$ \\
 &  & $^{22}$Na & $0.8\pm0.3$ & $0.56\pm0.06$ \\
 Cosmogenic &  & $^{121}$Te & & $2.38\pm0.57$ \\ 
 & & $^{123m}$Te & & $0.94\pm0.48$ \\ 
 & & $^{125m}$Te & & $0.25\pm0.41$ \\ \hline
\end{tabular} 
\medskip \\
\begin{tabular}{ccccc}
 \multicolumn{5}{c}{(b)} \\ \hline
 & & & \multicolumn{2}{c}{Activities} \\
 & & & \multicolumn{2}{c}{[mBq/PMT]} \\ \cline{4-5}
 Background sources & Group & Isotopes & Measured & Fitted \\ \hline
 & 1 & $^{238}$U & $25\pm5$ & $22.5\pm1.3$  \\
 & 2 & $^{234}$U & & $22.5\pm3.4$ \\ 
 & 3 & $^{230}$Th & & $22.5\pm3.3$ \\ 
 & 4 & $^{226}$Ra & & $150\pm10$ \\
 NaI-005 PMTs & 5 & $^{210}$Pb & & $22.5\pm2.7$ \\
 & 6 & $^{232}$Th & $12\pm5$ & $10.8\pm1.8$  \\ 
 & 7 & $^{228}$Ra & & $18\pm11$ \\
 & 8 & $^{228}$Th & & $14.4\pm2.5$ \\
 & 9 & $^{40}$K & $58\pm5$ & $63.8\pm10.0$ \\ \hline
 & 1 & $^{238}$U & $78.2\pm4.2$ & $70.4\pm10.6$ \\
 & 2 & $^{234}$U & & $70.5\pm8.0$ \\ 
 & 3 & $^{230}$Th & & $70.4\pm7.4$ \\ 
 & 4 & $^{226}$Ra & & $156.4\pm22.2$ \\
 CsI(Tl) PMTs & 5 & $^{210}$Pb & & $82.3\pm10.8$ \\
 & 6 & $^{232}$Th & $25.5\pm4.4$ & $23.0\pm3.4$ \\ 
 & 7 & $^{228}$Ra & & $140.3\pm23.1$ \\
 & 8 & $^{228}$Th & & $140.3\pm25.5$ \\
 & 9 & $^{40}$K & $504\pm72$ & $2772\pm196$ \\ \hline
 \end{tabular}
 \end{center}
\end{table*}

We simulated full decay chains of $^{238}$U, $^{232}$Th, and $^{40}$K from NaI(Tl) crystals and 26 PMTs that are grouped into three sets of nine background spectra and additionally we simulated five cosmogenic isotopes and bulk and surface $^{210}$Pb, including their progenies. 
Using all of the simulated background spectra, we fit the model to the measured data for both single-hit events and multiple-hit events of NaI-005, in the 2 to 1510-keV fitting range, to determine the unknown background fractions;  we used a maximum likelihood fit using Poisson statistics~\cite{fit-method}.
In the fitting, nine of the internal background groups were constrained by the measured activity errors and the other groups for external backgrounds were treated as floating and/or constrained parameters. All the cosmogenic background spectra and bulk/surface Pb210 spectra were considered as floating parameters in the fit.
The fitted results of activities of background sources in NaI-005 and PMTs are listed in Table~\ref{fit-result}.

Fig.~\ref{result} shows the fitted results for all the simulated background spectra plotted as various lines with different colors and styles for both single-hit events (a) and multiple-hit events (b) in the 0.5 to 2000-keV energy region. 
The overall energy spectrum summed over all simulations (solid red line) is well matched to the data (open black circles) not only for single-hit events but also for multiple-hit events. 
However, there are shown discrepancies between measurements and simulations for high energies above $\sim$1600~keV. 
It is because that the pulse shape above $\sim$1600~keV from the R12669SEL PMT is deformed and a significant fraction of the photoelectrons do not reach to the anode. Accordingly, the integrated charge signals of energetic events do not scale linearly with energy as one can see in Fig.~\ref{result}. It is also the reason why we chose the fitting range up to 1510-keV.

\begin{figure}[!htb]
\begin{center}
\includegraphics[width=0.5\textwidth]{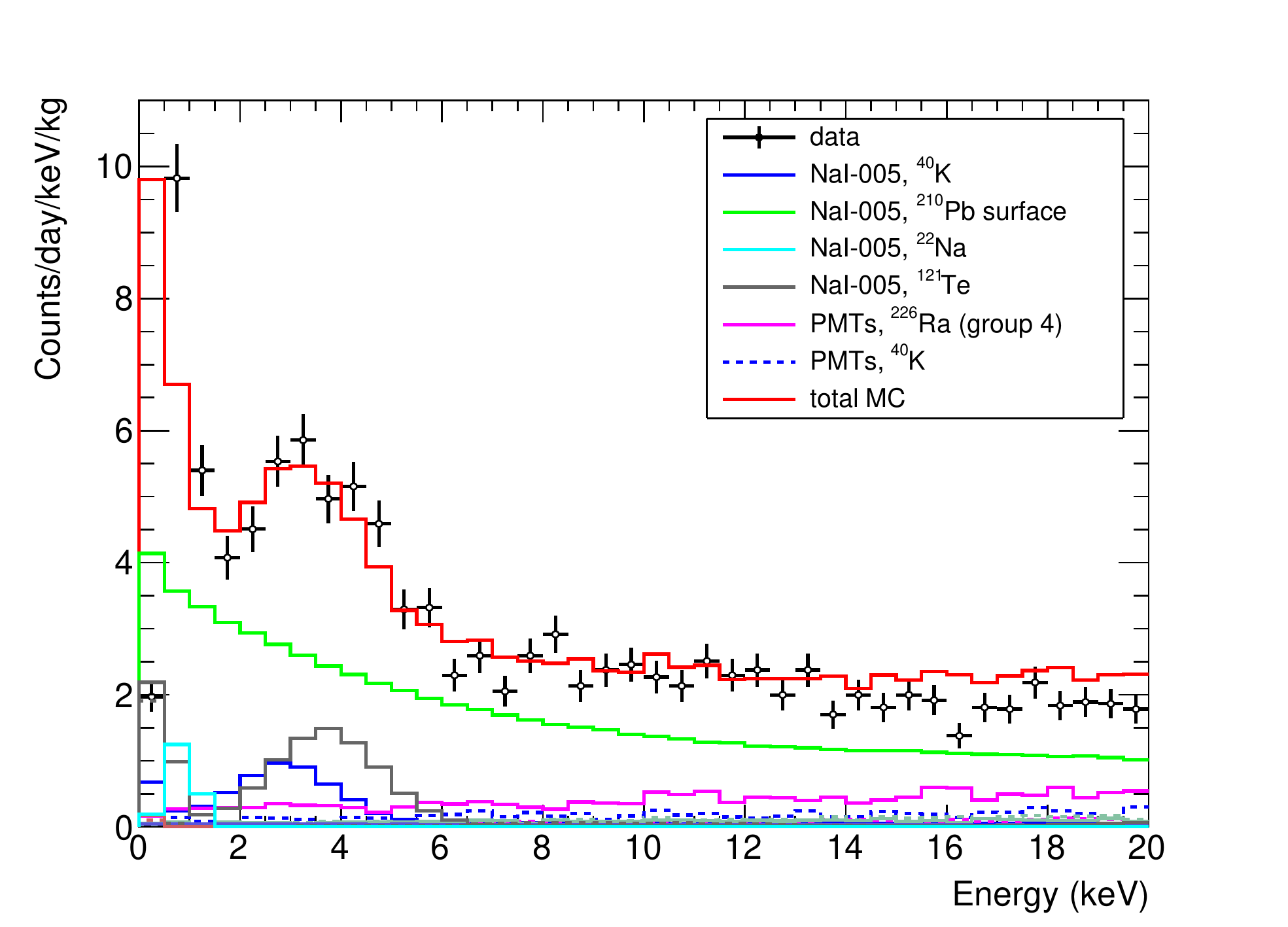}
\caption{Comparison of data and Monte Carlo simulation of the low-energy background spectra.}
\label{fit-result-lowenergyfig}
\end{center}
\end{figure}

The measured energy spectrum is compared with all the simulated spectra in the low-energy region below 20~keV in Fig.~\ref{fit-result-lowenergyfig}.
As listed in Table~\ref{fit-result-lowenergytab},
the main backgrounds in the 2  to 6-keV energy region are from surface $^{210}$Pb and $^{40}$K contaminations inside the crystal, the contributions of which are found to be 2.4 and 0.5 dru, respectively. 
Remnants of cosmogenic activation of  
Te (0.95 dru) still persist but will quickly be reduced with a lifetime of $<100$ days. 
We also determined the external backgrounds
, which amounted to 0.59 dru.    
In the low energy region below 0.5~keV the data is suppressed by the requirement for low energy noise rejection and there is shown less number of events in the data. 

\begin{table}[!htb]
\begin{center}
\caption{Simulated background events of NaI-005 in dru unit [/day/keV/kg] in the (2--6) keV energy interval.}
\label{fit-result-lowenergytab}
\begin{tabular}{cc} \hline
 &  Single-hit events \\
 Background sources & [/day/keV/kg] \\
 & Energy [keV] 2-6 \\ \hline
 $^{210}$Pb surface & $2.402\pm0.004$ \\
 $^{40}$K & $0.509\pm0.003$ \\
 Cosmogenic & $0.954\pm0.005$ \\
 External & $0.590\pm0.027$ \\
 Others & $0.074\pm0.001$ \\ \hline
 Total & $4.493\pm0.028$ \\ \hline
 \end{tabular}
 \end{center}
\end{table}

\section{Conclusion} 
\label{conclusion}
We have studied NaI(Tl) crystal backgrounds based on Monte Carlo simulation using the Geant4 toolkit. 
The studies show that all the simulated background spectra, normalized by measured activities and fitted fractions, describe the data well not only for single-hit events but also for multiple-hit events.
According to the comparison between the Monte Carlo simulation and the data in the low-energy spectra, the background is found to be dominated by $^{210}$Pb, mainly surface $^{210}$Pb, which is due to $^{222}$Rn exposure during crystal growing and/or handling procedures, and $^{40}$K within the NaI(Tl) crystal, the background level of which is consistent with ANAIS's expectation evaluated by assuming the activities of their characterized NaI(Tl) crystals~\cite{amare16}. 

External background contributions are expected to be tagged by an active veto detector with a liquid scintillator (LS) surrounding crystals in the new detector design.  
We already observed such a reduction, $0.76\pm0.04$ in the 6 to 20-keV energy region, with a prototype active veto system using an LS~\cite{kims-nai-prototype}. 
An LS veto detector can also reduce the contribution of the internal $^{40}$K background, by tagging 3-keV X-ray events with required conditions for the LS veto signal, by a factor of 2 with the optimized thickness of the LS in the new detector design~\cite{kims-nai-prototype}.
In addition, we are studying the suppression of the $^{210}$Pb crystal-surface background to achieve a background level as low as bulk $^{210}$Pb, which contributes 0.04~dru in the 2  to 6-keV energy region. Moreover, improving the purity of NaI(Tl) crystals with small concentrations of $^{210}$Pb, $\sim$60~$\mu$Bq/kg, is possible, as reported in reference to KamLAND-PICO~\cite{kamland-pico}. 

As a result, we are expecting that we can reach a background level of $<0.5$~dru in the 2 to 6-keV energy region. 

\section*{Acknowledgments}
This research was funded by the Institute for Basic Science (Korea) under project code IBS--R016--A1.

%
%

\end{document}